\documentstyle[aps,prb,twocolumn]{revtex}
\input{psfig}
\begin{document}
\draft
\title{Distribution of transmitted charge through a double-barrier junction}

\author{M. J. M. de Jong}
\address{ 
Philips Research Laboratories,
5656 AA  Eindhoven,
The Netherlands}
\date{Submitted May 6, 1996,  {\tt cond-mat/9605034}}

\maketitle

\begin{abstract}
The distribution function of transmitted charge through a double-barrier
junction is studied at zero temperature and at small applied voltage.
Both a semiclassical model, in which the transport is described by jump
rates, and a quantum mechanical model, which averages over resonant and
non-resonant states, are used to determine the characteristic function
of the transmitted electrons.
It is demonstrated that for large times the logarithm of the characteristic
function is equal within the two approaches.
The charge distribution 
is in between a Gaussian and a Poissonian distribution
if both barriers have equal height and
reduces to a Poissonian if one barrier is much higher than the other.

\end{abstract}


\section{Introduction}
\label{s1}

The nature of the current flow at low temperatures
through mesoscopic structures has
received a lot of attention during the last years.
After initial focus on the conductance, which measures the average
number of electrons transmitted in time, 
there has been an increasing interest in the noise power, 
a measure for the variance of the transmitted charge.
At zero temperature, these current fluctuations are due to the discreteness
of the electron charge.
It has been found theoretically,
that the zero-frequency shot-noise power $P$ can be
suppressed below its classical value 
characteristic for uncorrelated electron transport,
$P_{\rm Poisson}\equiv 2 e I$, with
$I$ the average current.
This suppression is due to correlated electron transmission imposed by
the Pauli principle \cite{khl87,les89,yur90,but90,mar92}.
Consequently, it has been shown that in a double-barrier junction
$P$ can be suppressed down to $\case{1}{2} P_{\rm Poisson}$
\cite{che91,dav92,her92,kor92,levy93,hun93,han93}, depending
on the relative height of the barriers.
For a metallic, diffusive conductor various calculations
yield that $P=\case{1}{3} P_{\rm Poisson}$ 
\cite{b&b92,nag92,jon92,naz94,alt94}.
The shot-noise suppression in these two systems has been observed 
experimentally \cite{li90,liu95,bir95,lie94,ste95}.

Recently, Levitov and Lesovik have gone one step further
by studying the full {\em distribution function} of charge
transmitted through a mesoscopic conductor \cite{lev93}.
This function gives the probability that a certain number of electrons
are transmitted during a given time interval.
Their quantum mechanical analysis demonstrates, that the attempts to transmit 
electrons are periodic in time, yielding a binomial distribution of transmitted
electrons. On the basis of this result, Lee, Levitov, and Yakovets have 
calculated the charge distribution function for transport
through a metallic, diffusive conductor
\cite{lee95}.

In this paper, we derive the complete distribution of transmitted charge
through a double-barrier junction, by two different methods:
Firstly, we follow
a semiclassical approach, in which phase is neglected 
but the Pauli principle is accounted for. Here, the electron transport is
described by classical jump rates.
Secondly, we take
a quantum mechanical approach, where we average the result of 
Ref.\ \onlinecite{lev93} over the distribution of
transmission probabilities through the double-barrier system.
For computational convenience, 
our analysis is carried out for a single-channel
conductor in the absence of spin degeneracy. However, the results can
directly be generalized to a multi-channel conductor.
Furthermore, we neglect charging effects, we assume small applied voltage
as well as zero temperature,
and we restrict ourselves to high tunnel barriers.

The linear-response conductance $G$ of
a double-barrier junction is given by
\begin{equation}
G=\frac{e^2}{h} \frac{ \Gamma_1 \Gamma_2}{\Gamma_1 + \Gamma_2}
\: ,
\label{eii1}
\end{equation}
with $\Gamma_i\ll 1$ the transmission probability through  barrier $i=1,2$.
Interestingly, the two approaches to derive Eq.\ (\ref{eii1})
are of a completely different nature \cite{but88,dav93}.
The semiclassical derivation
consists essentially in the addition of the resistances of both junctions,
whereas the quantum mechanical derivation involves an average
over resonant and non-resonant states.
Physically, this averaging may correspond either to an applied voltage larger
than the width of the resonance or to a summation over the modes
in a multi-channel conductor if the distance between the barriers is larger
than the Fermi wave length.

The role of the presence of phase coherence
on the fluctuations in the current is still
an intriguing issue \cite{shi92,lan93,dav95}. 
For example, the one-third suppression of the shot
noise in a metallic, diffusive conductor was originally surmised
to be of quantum mechanical origin \cite{b&b92}.
However, later derivations through a semiclassical approach
yielded a suppression by one-third as well \cite{nag92,jon95,jon96}.
With respect to the shot-noise power in the double-barrier junction,
a quantum mechanical theory
by Chen and Ting \cite{che91} and a semiclassical theory by
Davies {\em et al}.\ \cite{dav92} give identical results, namely
\begin{equation}
P=\frac{ {\Gamma_1}^2 + {\Gamma_2}^2}{(\Gamma_1 + \Gamma_2)^2}
P_{\rm Poisson} \: .
\label{eii2}
\end{equation}
An additional aim of the present paper 
is to check to which extent this insensitivity to the presence of phase 
coherence applies also
for the complete distribution of transmitted charge.

The quantity of interest is $P_n(t)$, denoting the probability that exactly
$n$ electrons have been transmitted during a time interval $t$.
An alternative way to describe this distribution is
through its characteristic function $\chi(\lambda,t)$.
They are mutually related according to \cite{Crame}
\begin{mathletters}
\label{ei1}
\begin{eqnarray}
\chi(\lambda,t)&=&\sum \limits_{n=0}^\infty P_n(t) e^{i n \lambda} \: ,
\label{ei1a} \\
P_n(t) &=& \frac{1}{2 \pi} \int \limits_{-\pi}^\pi d \lambda \,
e^{-i n \lambda} \chi(\lambda,t) \: .
\label{ei1b}
\end{eqnarray}
\end{mathletters}%
It is often more convenient to determine $\chi(\lambda,t)$ instead of
$P_n(t)$.
Normalization requires that $\chi(0,t)=1$. Furthermore, it follows from Eq.\
(\ref{ei1a}) that the average number of electrons transmitted during a time
$t$ is given by
\begin{equation}
\overline{n(t)} = \sum  \limits_{n=0}^\infty  n P_n(t) =
\lim \limits_{\lambda \rightarrow 0}
\frac{\partial}{i \partial \lambda} \chi(\lambda,t) \: .
\label{ei2}
\end{equation}
More generally, one can express the $k$th moment $\mu_k(t)$ of the distribution
according to 
\begin{equation}
\mu_k(t) \equiv \overline{n^k(t)} = \lim \limits_{\lambda \rightarrow 0}
\left( \frac{\partial}{i \partial \lambda} \right)^k \chi(\lambda,t) \: .
\label{ei3}
\end{equation}
We note that the average current is simply $I=e \mu_1(t)/t$, whereas
the noise power is proportional to the variance of the number of transmitted
electrons
$P=2 e^2 \lim_{t \rightarrow \infty} \mbox{var} \, n(t) / t =
2 e^2 \lim_{t \rightarrow \infty} [\mu_2(t) - {\mu_1}^2(t)]/t$.
The logarithm of the characteristic function can be expanded 
as follows
\begin{equation}
\ln \chi(\lambda,t) = \sum \limits_{k=1}^\infty \frac{(i \lambda)^k}{k!}
\kappa_k(t) \: ,
\label{ei4}
\end{equation}
with $\kappa_k(t)$ the $k$th 
cumulant of the distribution.
The moments and the cumulants have a direct polynomial relation
\cite{Crame}, for example $\kappa_1(t)=\mu_1(t)$,
$\kappa_2(t)=\mu_2(t) - {\mu_1}^2(t)$. 

The quantum mechanical analysis by Levitov and Lesovik \cite{lev93}
has yielded 
the characteristic function of the charge through a single-channel
conductor at zero temperature and at small voltage $V$,
\begin{equation}
\chi(\lambda,t)=[ (e^{i \lambda}-1)T + 1 ]^{eVt/h} \: ,
\label{ei5}
\end{equation}
with $T$ the transmission probability at the Fermi level through the
conductor.
 From Eqs.\ (\ref{ei3}) and (\ref{ei5}) one can immediately
derive the Landauer formula for the conductance $G$
and the formula for the shot-noise power \cite{khl87,les89} 
\begin{eqnarray}
G&=& \frac{e^2}{h} T \: ,
\label{ei6} \\
P&=&\frac{2 e^3 V}{h} T(1-T) \: .
\label{ei7}
\end{eqnarray}
Note that Eq.\ (\ref{ei7}) is only valid in the
phase-coherent regime, whereas in the absence of phase coherence
$P$ is given by a different equation.\cite{jon95}
In Sec.\ \ref{s2}, it is demonstrated how the charge distribution
through a single barrier can be derived in a semiclassical approach.
Sec.\ \ref{s3} repeats this analysis for the double-barrier junction.
The quantum mechanical calculation is given in Sec.\ \ref{s4},
after which we conclude in Sec.\ \ref{s5}.

\section{Simple example: single-barrier junction}
\label{s2}

Let us illustrate our approach, by calculating the distribution of
transmitted charge through a single-channel, single-barrier junction,
with a transmission probability $\Gamma \ll 1$.
The average current through the barrier 
$I=e^2 V \Gamma /h \equiv e \gamma$,
with $\gamma= e V \Gamma/h$ the tunnel rate through the barrier.
The probability $P_n(t)$ that $n$ electrons have been transmitted
in a time $t$ obeys the master equation
\begin{mathletters}
\label{e2}
\begin{eqnarray}
\frac{d P_0(t)}{dt}&=& - \gamma P_{0}(t) \: ,
\label{e2a} \\
\frac{d P_n(t)}{dt} &=& \gamma P_{n-1}(t) - \gamma P_{n}(t) \: ,
\quad \mbox{if} \ n \geq 1 \: ,
\label{e2b}
\end{eqnarray}
\end{mathletters}%
with the initial condition $P_n(0) = \delta_{0,n}$.
Eq.\ (\ref{e2}) can be solved straightforwardly by various means.
Here, we adopt an approach, which appears to be useful for the
double-barrier junction.
The solution of Eq.\ (\ref{e2a}) is $P_0(t)=\exp(- \gamma t)$.
We define the waiting-time distribution $\psi(t)\equiv - dP_0(t)/dt$,
denoting the probability density that an electron is transmitted
immediately after having waited a time $t$,
\begin{equation}
\psi(t)=\gamma e^{- \gamma t} \: .
\label{e2c}
\end{equation}
We write
\begin{equation}
P_n(t) = G_n(t) - G_{n+1}(t) \: ,
\label{e3}
\end{equation}
where $G_n(t)$ denotes the probability that $n$ or more electrons are
transmitted during a time $t$.
It can be calculated according to
\begin{eqnarray}
G_n(t)&=&\int \limits_0^t dt_1 \int \limits_{t_1}^t dt_2 \cdots
\int \limits_{t_{n-1}}^t dt_n 
\nonumber \\ &\times&
\psi(t_1) \psi(t_2-t_1) \cdots
\psi(t_n - t_{n-1}) \: .
\label{e4}
\end{eqnarray}
Since this is a convolution, one has for the Laplace transform
\begin{equation}
\widetilde{G}_n(s)\equiv \int \limits_0^\infty dt \, e^{-st} G_n(t)
=\frac{1}{s} [ \widetilde{\psi}(s) ]^n \: ,
\label{e5}
\end{equation}
where the Laplace transform of Eq.\ (\ref{e2c}) is given by
\begin{equation}
\widetilde{\psi}(s) = \frac{\gamma}{s+\gamma} \: .
\label{e6}
\end{equation}
 From Eqs.\ (\ref{e3}), (\ref{e5}), and (\ref{e6}), we find
\begin{equation}
\widetilde{P}_n(s)=\frac{{\gamma}^n}{(s+\gamma)^{n+1}} \: ,
\label{e7}
\end{equation}
yielding for the distribution in time
\begin{equation}
P_n(t)=\frac{(\gamma t)^n}{n!} e^{-\gamma t} \: .
\label{e8}
\end{equation}
This is the Poisson distribution, as one would expect for the 
uncorrelated electron transfers through the barrier.
The characteristic function is given by
\begin{equation}
\chi(\lambda,t)=\exp[ \gamma t (e^{i \lambda} -1) ] \: .
\label{e9}
\end{equation}
Indeed, in the limit $T=\Gamma \rightarrow 0$, Eqs.\ (\ref{ei5}) and
(\ref{e9}) coincide.

\section{Classical approach}
\label{s3}

We now study the double-barrier junction.
The tunnel rate through barrier $i=1,2$ is $\gamma_i = e V \Gamma_i/h$.
Due to the Pauli principle, the number of electrons in the
double-barrier junction can be either 0 or 1.

\begin{figure}
\centerline{\psfig{file=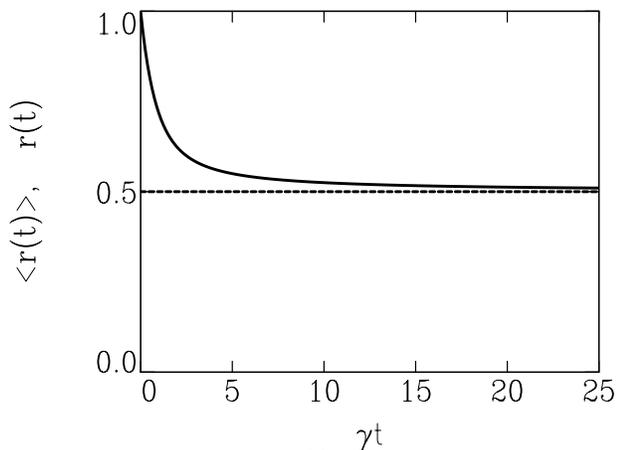,width=8cm}}
\caption{
The Fano factor $r(t)$, giving the ratio of the variance to the average
number of transmitted electrons, versus time $t$
for a symmetric double-barrier junction
with tunnel rates $\gamma_1=\gamma_2=\gamma$.
The solid line gives the result (\ref{e18}) of the semiclassical analysis
 and 
the dashed line the quantum result $\langle r(t) \rangle$,
according to Eq.\ (\ref{e24d}).
}
\label{f1}
\end{figure}

Our analysis is similar to the single-barrier case. However, one now has to
take into account two
possible initial conditions at $t=0$: either 0 electrons in the junction
--- a situation with probability $\gamma_2/(\gamma_1+\gamma_2)$ ---
or 1 electron --- probability $\gamma_1/(\gamma_1+\gamma_2)$.
The distribution of transmitted charge is thus given by
\begin{equation}
P_n(t) = \frac{\gamma_2}{\gamma_1+\gamma_2} P_n^{(0)}(t) + 
\frac{\gamma_1}{\gamma_1+\gamma_2} P_n^{(1)}(t) \: ,
\label{e10}
\end{equation}
where $P_n^{(j)}(t)$ starts from $j$ 
electrons in the junction at $t=0$.
The probability that at least $n$ electrons have been transmitted 
can be expressed as
\begin{mathletters}
\label{e11}
\begin{eqnarray}
\widetilde{G}_n^{(0)}(s)&=&\frac{1}{s} 
[ \widetilde{\psi}_1(s) \widetilde{\psi}_2(s) ]^n \: ,
\label{e11a} \\
\widetilde{G}_n^{(1)}(s)&=&\frac{1}{s} \widetilde{\psi}_2(s)
[ \widetilde{\psi}_1(s) \widetilde{\psi}_2(s) ]^{n-1} \: ,
\label{e11b}
\end{eqnarray}
\end{mathletters}%
with $\widetilde{\psi}_i(s)=\gamma_i/(s+\gamma_i)$.
 From Eqs.\ (\ref{e3}), (\ref{e10}), and (\ref{e11}) we obtain 
\begin{mathletters}
\label{e12}
\begin{eqnarray}
\widetilde{P}_0(s)&=& 
\frac{ s (\gamma_1 + \gamma_2) + 
{\gamma_1}^2 + {\gamma_2}^2 + \gamma_1 \gamma_2}
{ (\gamma_1 + \gamma_2) (s + \gamma_1)(s+\gamma_2)} \: ,
\label{e12a} \\
\widetilde{P}_n(s)&=& 
\frac{ (\gamma_1 \gamma_2)^n (s + \gamma_1+\gamma_2)^2}
{(\gamma_1+\gamma_2)(s + \gamma_1)^{n+1} (s+\gamma_2)^{n+1}}
\: , 
\nonumber \\ &&
\mbox{if} \ n \geq 1 \: .
\label{e12b}
\end{eqnarray}
\end{mathletters}%
The distribution function in time can be obtained through the inverse
Laplace transform. In general, it yields a rather cumbersome expression.
However, for the case of a symmetric double-barrier junction,
$\gamma_1=\gamma_2 \equiv \gamma$, one has
\begin{mathletters}
\label{e13}
\begin{eqnarray}
P_0(t)&=&\left(1+ \frac{\gamma t}{2}\right) e^{-\gamma t} \: ,
\\
P_n(t)&=&\left[ 
\frac{(\gamma t)^{2n-1}}{2 (2n-1)!} +
\frac{(\gamma t)^{2n}}{(2n)!} +
\frac{(\gamma t)^{2n+1}}{2 (2n+1)!}
\right] e^{- \gamma t} \: ,
\nonumber \\ &&
\mbox{if} \ n \geq 1 \: .
\end{eqnarray}
\end{mathletters}%
For arbitrary $\gamma_1$ and $\gamma_2$, we can evaluate from Eq.\
(\ref{e12}) the Laplace transform of the characteristic function
\begin{equation}
\widetilde{\chi}(\lambda,s) = \frac{1}{\gamma_1+\gamma_2}
\left[ \frac{(s+\gamma_1+\gamma_2)^2}{(s+\gamma_1)(s+\gamma_2) -
e^{i\lambda} \gamma_1 \gamma_2} - 1 \right] \: ,
\label{e14}
\end{equation}
yielding for the characteristic function in time
\begin{eqnarray}
\chi(\lambda,t)&=& \exp [-\case{1}{2}(\gamma_1+\gamma_2) t ]
\left\{ \cosh [\case{1}{2} \beta(\lambda) t ] 
\vphantom{ \frac{e^{i \lambda}}{\beta} }
\right.
\nonumber \\
&+& \left. 
\left(
\frac{\beta(\lambda)}{\gamma_1+\gamma_2}
-\frac{2 \gamma_1 \gamma_2(e^{i \lambda} -1)}
{(\gamma_1+\gamma_2)\beta(\lambda)}\right)
\sinh [\case{1}{2} \beta(\lambda) t ] \right\}
\: ,
\nonumber \\ &&
\label{e15}
\end{eqnarray}
with $\beta(\lambda)=\sqrt{(\gamma_1+\gamma_2)^2 + 
4\gamma_1 \gamma_2 (e^{i\lambda}-1)} $.
Eq.\ (\ref{e15}) is the central result of our semiclassical analysis.
Let us evaluate from Eqs.\ (\ref{ei3}) and (\ref{e15})
the first two moments of the charge distribution.
The average number of electrons transmitted during a time $t$ is
\begin{equation}
\overline{n(t)}=
\mu_1(t)=\frac{\gamma_1 \gamma_2}{\gamma_1+\gamma_2} t \: ,
\label{e16}
\end{equation}
in agreement with Eq.\ (\ref{eii1}).
For the second moment we find
\begin{eqnarray}
\overline{n^2(t)}=\mu_2(t)&=&
\frac{(\gamma_1 \gamma_2)^2}{(\gamma_1+\gamma_2)^2} t^2 +
\frac{\gamma_1 \gamma_2({\gamma_1}^2 + {\gamma_2}^2)}{(\gamma_1+\gamma_2)^3} t
\nonumber \\
&+&
\frac{2 (\gamma_1 \gamma_2)^2}{(\gamma_1+\gamma_2)^4}
\{ 1 - \exp[-(\gamma_1+\gamma_2) t] \} \: .
\label{e17}
\end{eqnarray}
The Fano factor, defined as the ratio of the variance to the average
number of transmitted electrons,
$r(t)\equiv [\mu_2(t)-{\mu_1}^2(t)]/\mu_1(t)$, 
follows from Eqs.\ (\ref{e16}) and (\ref{e17}),
\begin{equation}
r(t)=\frac{{\gamma_1}^2 + {\gamma_2}^2}{(\gamma_1+\gamma_2)^2}
+ \frac{2 \gamma_1 \gamma_2 \left( 1 - e^{-(\gamma_1+\gamma_2) t}  \right) }
{(\gamma_1+\gamma_2)^3 \, t} \: .
\label{e18}
\end{equation}
The Fano factor
gives the relative
magnitude of the current fluctuations.
Indeed, for large $t$, $r(t)$ yields the shot-noise suppression
according to Eqs.\ (\ref{eii1}) and (\ref{eii2}). 
In Fig.\ \ref{f1} we have plotted 
$r(t)$, for $\gamma_1=\gamma_2\equiv\gamma$.
We find that $r(t)$ goes from $1$ at small $t$, indicative for uncorrelated
electron transmission, to $\case{1}{2}$ at large $t$, indicative
for a more correlated electron transmission.
It is already within one percent of its final value at $\gamma t=50$,
which corresponds to an average number of 25 transmitted electrons.

The cumulants of the charge distribution can be determined from
the logarithm of Eq.\ (\ref{e15})
\begin{eqnarray}
\ln \chi(\lambda,t) &=&
\case{1}{2} [ \beta(\lambda) - (\gamma_1 + \gamma_2) ] t
\nonumber \\
&+& \ln \left( 1 + e^{-\beta(\lambda)t} \right) - \ln 2
\nonumber \\
&+& \ln \left[ 1 + \left( \frac{\beta(\lambda)}{\gamma_1 + \gamma_2}
- \frac{2 \gamma_1 \gamma_2 (e^{i \lambda} - 1)}
{(\gamma_1 + \gamma_2) \beta(\lambda)} \right) \right.
\nonumber \\
&& \times \left. 
\vphantom{ \frac{\gamma_2 (e^{i \lambda}}{\beta(\lambda)}}
\tanh [ \case{1}{2} \beta(\lambda) t ] \, \right]
\: .
\label{e19}
\end{eqnarray}
In general, it is cumbersome to derive the cumulants from this result.
However, for large $t$, only the first term remains of importance.

\section{Quantum mechanical approach}
\label{s4}

\begin{figure}
\centerline{\psfig{file=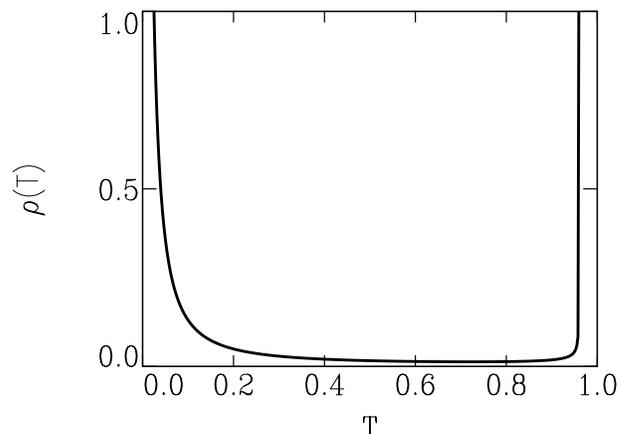,width=8cm}}
\caption{
The distribution of transmission probabilities through a double-barrier
junction in a quantum mechanical model, according to
Eq.\ (\ref{e21}), for $\Gamma_1=0.02$ and $\Gamma_2=0.03$.
}
\label{f2}
\end{figure}

Whereas in a semiclassical picture, the transmission probability through
the double-barrier junction is just a constant, 
in a quantum mechanical approach,
the transmission probability $T$ 
varies according to a Fabry-Perot type of formula
\begin{equation}
T=\frac{\Gamma_1 \Gamma_2}{1 - 2 \sqrt{(1-\Gamma_1)(1-\Gamma_2)} \cos\phi
+ (1-\Gamma_1)(1-\Gamma_2) }
\: ,
\label{e20}
\end{equation}
where $\phi$ is the phase accumulated in one round trip between the
barriers.
The distribution function $\rho(T)$ of the transmission probabilities
through the system can be obtained from the assumption
that $\phi$ is uniformly distributed between 0 and $2 \pi$.\cite{mel94}
In the limit $\Gamma_1, \Gamma_2 \ll 1$ this implies 
\begin{equation}
\rho(T) = \frac{\Gamma_1 \Gamma_2}{\pi (\Gamma_1+\Gamma_2)} \,
\frac{1}{\sqrt{T^3(T_1 - T)} } \: ,
\label{e21}
\end{equation}
if $T \in [ T_0, T_1 ]$, 
and $\rho(T)=0$ otherwise,
with $T_0=4 \Gamma_1 \Gamma_2/ [
(\Gamma_1+\Gamma_2)^2 + 4 \pi^2 ]$ and
$T_1=4 \Gamma_1 \Gamma_2/(\Gamma_1+\Gamma_2)^2$.
The distribution function (\ref{e21}) is plotted in Fig.\ \ref{f2}.
Similar to a metallic, diffusive conductor \cite{b&b92},
this distribution is bimodal, in the sense that the transmission
probabilities are either close to $T_0 \approx 0$ or close to $T_1$. 

The ensemble average of a quantity $a(T)$
over all possible transmission probabilities is given by
$\langle a \rangle = \int_{T_0}^{T_1} dT \rho(T) a(T)$.
It is convenient 
to switch variables from $T$ to $\nu$ with
$T=T_1/(1+\nu^2)$, so that
$\rho(\nu)=\rho_0\equiv(\Gamma_1+\Gamma_2)/2 \pi$ is uniform over the range
$[0,\nu_{max}]$. 
In practice, the upper limit $\nu_{max}$ can be often replaced
by infinity.
The ensemble average for the $m$th power ($m\geq 1$) 
of the transmission probability is given by
\begin{eqnarray}
\langle T^m \rangle &=& \rho_0 \int \limits_0^\infty d\nu \, 
\left( \frac{T_1}{ 1 + \nu^2}\right)^m
\nonumber \\
&=& 
\frac{(2m-2)!}{[(m-1)!]^2} \, 
\frac{(\Gamma_1 \Gamma_2)^m}{(\Gamma_1 + \Gamma_2)^{2m-1}}
\: .
\label{e22}
\end{eqnarray}
Substituting this result into Eqs.\ (\ref{ei6}) and (\ref{ei7}), we 
recover for
$\langle G \rangle$ and $\langle P \rangle$ the expressions given by
Eqs.\ (\ref{eii1}) and (\ref{eii2}).

In order to obtain the ensemble average of all the cumulants of the
distribution function, we 
average the logarithm of the characteristic function \cite{lee95}.
For the double-barrier junction, we obtain from Eq.\ (\ref{ei5})
\begin{eqnarray}
\langle \ln \chi(\lambda,t) \rangle
&=& 
\rho_0 \frac{e V t}{h}
\int \limits_0^\infty d\nu \, 
\ln \left[ \frac{(e^{i\lambda}-1)T_1}{1+\nu^2} +1 \right]
\nonumber \\
&=& \case{1}{2} [ \beta(\lambda) - (\gamma_1 + \gamma_2) ] t
\: ,
\label{e23}
\end{eqnarray}
with $\beta(\lambda)=\sqrt{(\gamma_1+\gamma_2)^2 + 
4\gamma_1 \gamma_2 (e^{i\lambda}-1)}$.
This is the key result of the quantum mechanical evaluation.
Using Eq.\ (\ref{ei4}), we find for the ensemble average of the
first three cumulants
\begin{mathletters}
\label{e24}
\begin{eqnarray}
\langle \kappa_1(t) \rangle &=& 
\frac{\gamma_1 \gamma_2}{\gamma_1+\gamma_2} \, t
\: , \label{e24a} \\
\langle \kappa_2(t) \rangle &=&  
\frac{\gamma_1 \gamma_2({\gamma_1}^2+{\gamma_2}^2)}{(\gamma_1+\gamma_2)^3}
\, t
\: , \label{e24b} \\
\langle \kappa_3(t) \rangle &=& 
\frac{\gamma_1 \gamma_2}
{(\gamma_1+\gamma_2)^5} \, t
\nonumber \\ &\times&
( {\gamma_1}^4 - 2{\gamma_1}^3 \gamma_2 + 6 {\gamma_1}^2 {\gamma_2}^2
- 2\gamma_1 {\gamma_2}^3 + {\gamma_2}^4 )
\: . 
\nonumber \\ &&
\label{e24c}
\end{eqnarray}
\end{mathletters}%
Since $\langle \ln \chi(\lambda,t) \rangle$ is proportional to $t$, 
all the cumulants are linear in $t$ as well.
This implies for the Fano factor
\begin{equation}
\langle r(t) \rangle \equiv 
\frac{\langle \kappa_2(t) \rangle}{\langle \kappa_1(t) \rangle}=
\frac{ {\gamma_1}^2+{\gamma_2}^2}{(\gamma_1 + \gamma_2)^2}
\: ,
\label{e24d}
\end{equation}
which
is constant in time and equal to the large-$t$ value of Eq.\ (\ref{e18})
(see Fig. \ref{f1}).
If $\gamma_1 \gg \gamma_2$ (or vice versa), Eq.\ (\ref{e23}) reduces to 
a Poissonian distribution, as expected.
For a symmetric double-barrier junction with
$\gamma_1=\gamma_2\equiv\gamma$, the expression
(\ref{e23}) simplifies considerably:
\begin{equation}
\langle \ln \chi(\lambda,t) \rangle =
\gamma t \left( e^{i \lambda/2} -1 \right)
\: .
\label{e25} 
\end{equation}
%Through Eq.\ (\ref{ei1b}) we can obtain an estimation for the
%transmitted charge distribution
%\begin{equation}
%\langle P_n(t) \rangle \simeq
%\left[ \frac{(\gamma t)^{2n}}{(2n)!} +
%\sum \limits_{m=0}^{\infty} 
%\frac{ (-1)^{m-n} (\gamma t)^{2m+1}}{\pi (m-n+\case{1}{2}) (2m+1)!}
%\right] e^{-\gamma t}
%\: .
%\label{e27}
%\end{equation}
%This result is not exact, since it is based on the transform
%of $\exp[ \langle \ln \chi \rangle ]$ instead of $\langle \chi \rangle$.
For the $k$th cumulant we find from Eqs.\ (\ref{ei4}) and (\ref{e25})
\begin{equation}
\langle \kappa_k(t) \rangle = \frac{\gamma t}{2^k} \: .
\label{e26}
\end{equation}
The charge distribution is thus somewhere between
Gaussian (where $\kappa_k=0$ for $k\geq 3$) and Poissonian
(where $\kappa_k =\gamma t$ for all $k$).

\section{Conclusions}
\label{s5}

Let us make the comparison
between the outcome of the two approaches, i.e., between
the semiclassical result $\ln \chi(\lambda,t)$ from
Eq.\ (\ref{e19}) and the quantum result 
$\langle \ln \chi(\lambda,t) \rangle$ given in Eq.\ (\ref{e23}),
where we have averaged over the distribution of transmission probabilities.
The semiclassical result yields a more complicated expression, however
the most important contribution, which is proportional to $t$, is
{\em precisely} equivalent to the result (\ref{e23}).
The other terms are either constant (do not depend on $t$) or vanish
exponentially with $t$.
Only on short time scales, corresponding to the transfer of a few
electrons, we see sizeable differences between both approaches.
We surmise that these differences at small $t$ are not due
to the neglect of the phase, but rather depend on the precise
way the reservoirs are modeled in both approaches.
The quantum mechanical derivation is based on Eq.\ (\ref{ei5}), which
applies for arbitrary transmission probability between 0 and 1.
Here, the number of electrons transmitted has a maximum
value $n(t)=eVt/h$.
Our semiclassical calculation assumes independent tunnel events and is
therefore only valid for small transmission probabilities.
However, the number of electrons which can be transmitted $n(t)$ is 
not bounded. 
Even though this seems not to be very important for the case of high tunnel
barriers, one may expect that it leads to differences on a small time
scale. 
Therefore, we just draw conclusions from the comparison at larger $t$.
Here we find that the two results
are equal, and that as a consequence, all the cumulants are also equal.
This demonstrates that the statistics of 
charge transport through a double-barrier junction
does not reveal whether phase coherence is present or absent.

This insensitivity to the presence of phase coherence, does not imply
that phase breaking is not of influence in a real experiment.
This depends on the physical process which destroys the phase coherence.
For example, using the method given in Ref.\ \onlinecite{jon96},
one can show that electron-electron scattering
inside the double-barrier system, in which both phase
coherence is destroyed and energy is redistributed among the electrons,
increases the shot noise above the value of Eq.\ (\ref{eii2}).
However, the analysis in the present paper demonstrates that
merely breaking the phase leaves the charge transport through the
system unaffected.
This is in contrast to the result of Ref.\ \onlinecite{dav95},
in which incoherence is modeled by adding random phases to the wave
function on each round trip. The authors find that 
this increases the shot noise,
so that we conclude that their model is not equivalent to just 
destroying the phase.

It might be interesting to determine 
the role of inelastic processes inside the tunnel barrier
on the charge distribution.
We think that these effects
can well be taken into account using the semiclassical analysis, whereas
a complete quantum mechanical derivation looks more complicated.
Another extension of the work described in this paper, would be to repeat
the semiclassical analysis for a metallic, diffusive conductor, and compare
the outcome with the quantum mechanical derivation of Ref.\ 
\onlinecite{lee95}.

In summary, we have derived the complete distribution of transmitted charge
through a double-barrier junction at zero temperature and at low voltage.
We have used a semiclassical approach on the basis of classical jump rates
as well as a quantum mechanical approach, in which the result of
Levitov and Lesovik for an arbitrary single-channel conductor \cite{lev93}
is averaged over the distribution of transmission
probabilities through the system.
Our results are in precise agreement with previous values for the
conductance and for the shot-noise power \cite{che91,dav92}.
Within both approaches, we have determined the logarithm of the characteristic
function, which become equivalent at large times.
It is found that for symmetric tunnel barriers, the charge distribution is
between a Gaussian and a Poissonian distribution.

\acknowledgements

I would like to thank C. W. J. Beenakker and L. F. Feiner for useful
comments.

\end{document}